\documentclass[preprint,12pt]{elsarticle}
\usepackage[super,compress]{cite}
\usepackage{graphicx}
\begin{document}
\markboth{HYWEL OWEN, JOSE ALONSO, RANALD MACKAY, DAVID HOLDER}
{BEAM DELIVERY FOR PROTON THERAPY}

%\catchline{}{}{}{}{}
%
%%%%%%%%%%%%%%%%%%%%%%%%%%%%%%%%%%%%%%%%%%%%%%%%%%%%%%%%%%%%%%%%%%%%

\title{TECHNOLOGIES FOR DELIVERY OF PROTON AND ION BEAMS FOR RADIOTHERAPY}

\author{HYWEL OWEN\footnote{}, DAVID HOLDER}

\address{School of Physics and Astronomy, Schuster Laboratory\\
University of Manchester/Cockcroft Institute\\
Manchester M13 9PL,
United Kingdom\\
hywel.owen@manchester.ac.uk,david.holder@cockcroft.ac.uk}

\author{JOSE ALONSO}

\address{Accelerator and Fusion Research Division\\
Lawrence Berkeley National Laboratory\\
1 Cyclotron Road, Berkeley\\
CA 94720, USA\\
jralonso@lbl.gov}

\author{RANALD MACKAY}

\address{Christie Medical Physics and Engineering\\
The Christie NHS Foundation Trust\\
Manchester\\
United Kingdom\\
ranald.mackay@christie.nhs.uk}

\maketitle

%\begin{history}
%\received{Day Month Year}
%\revised{Day Month Year}
%\end{history}

\begin{abstract}
\textbf{Recent developments for the delivery of proton and ion beam therapy have been significant, and a number of technological solutions now exist for the creation and utilisation of these particles for the treatment of cancer.
In this paper we review the historical development of particle accelerators used for external beam radiotherapy and discuss the more recent progress towards more capable and cost-effective sources of particles.}

%\keywords{Keyword1; keyword2; keyword3.}
\end{abstract}

%\ccode{PACS numbers:}

%\tableofcontents

\section{Introduction to External Beam Radiotherapy}	

Soon after the discovery of x-rays in by R\"{o}ntgen in 1895, the therapeutic properties of radiation were being explored.
Radiation was used initially as a treatment for a number of non-cancerous and cancerous conditions.
In modern times radiotherapy has been used almost exclusively in the curative and palliative treatment of cancer, often in combination with surgery and chemotherapy \cite{thariatpast:2013}.
Radiotherapy may be divided into two principal categories:
\begin{itemize}
  \item Brachytherapy -  the use of sealed and unsealed radioactive sources placed near to or within the tumour requiring treatment \cite{Norderhaug:2003ez,Koukourakis:2009fs};
  \item External beam therapy (originally termed teletherapy) - the use of an external beam of radiation, usually produced by a particle accelerator \cite{frya1948,Bortfeld:2006jd,greene97} but also by radioactive sources \cite{johns10001952,VanDyk:1996uh,Ravichandran:2009ii,williams00}.
\end{itemize}
Teletherapy beams are provided either directly or from a source of secondary radiation; for example x-ray generation from electrons striking a Bremsstrahlung target.
Fast neutrons, or more exotic particles such as pions, are also possible.
In addition there are treatment involving two steps, such as boron neutron-capture therapy (BNCT), in which epithermal neutrons are absorbed by $\,^{10}$B nuclei which have chemically bound to tumour cells, giving rise to alpha particle emission and hence a very local radiation dose.

The amount of energy deposited as radiation passes through a given depth of tissue is referred to as the linear energy transfer (LET).
X-rays and protons are considered to be low LET radiations, while other hadronic particle beams such as neutrons and carbon ions are considered to have high LET.
The predominant type of external beam radiotherapy is x-ray therapy, where a small (typically less than 2 metres) standing-wave or travelling-wave linear accelerator (linac) accelerates electrons to an energy of around 10~MeV typically.
When incident upon a transmission target these electrons produce Bremsstrahlung radiation.
This photon beam can then be collimated and flattened to produce a uniform beam of defined field size.
In the UK around 130,000 treatments a year are presently delivered involving 2.5 million attendances, more than half of which are for breast and prostate treatment \cite{radiotherapyengland}.

Although the linacs used for x-ray therapy are both relatively compact and produced in large numbers, and therefore relatively inexpensive compared to proton and other ion accelerators, the resultant radiation dose delivered by the photons within the patient is not ideal.
Whilst photons give a low surface dose - providing valuable skin sparing -  this dose rises rapidly to a maximum within the first $\approx$2.5 cm of tissue before falling with depth due to attenuation and the inverse-square law.
A single beam direction will therefore deliver lower dose at the depth of the tumour than it does to healthy tissue upstream and will unnecessarily irradiate healthy tissue downstream the tumour.
These inherent limitations may be partly overcome to give better conformation to the tumour of the delivered dose by:
\begin{itemize}
  \item Bringing beams onto the target volume from a number of directions;
  \item Defining the transverse shape each beam using a multi-leaf collimator (MLC);
  \item Varying the intensity of each beam through the technique known as intensity-modulated radiation therapy (IMRT).
\end{itemize}

Through these means modern x-ray radiotherapy can provide good conformation of the high-dose volume to the target, but inherent in the treatment is the irradiation of large amounts of healthy tissue with medium and low doses.
Good treatment planning seeks to optimise this tradeoff and to minimise dose to sensitive structures \cite{Webb:1997ui,Webb:2001wu,Mayles:2007ua,Barratt:2009tf,Hoskin:2012vm}.
There are presently around 265 linacs in clinical use in the UK and a funded programme to dramatically expand the use of IMRT \cite{radiotherapyengland}.
Three-dimensional computed tomography (CT) provides sufficient resolution of the electron density to x-rays \cite{Khoo:1997vs,Khoo:2006ea,Dawson:2010eq} given that the depth-dose curve for x-ray absorption is rather smooth.
This is supported by magnetic resonance imaging (MRI) to improve targeting, particularly in soft tissues.
Computer-aided optimisation of the treatment dose is used with a number of irradiation fields to spare tissues and organs at risk that are near to the tumour \cite{Webb:1997ui,Nutting:2000vy,Webb:2001wu,Garden:2004ht,Bortfeld:2006jd,Veldeman:2008gt,Staffurth:2010dh}.

\section{Particle Therapy}	

In contrast to x-ray therapy, radiotherapy with charged hadronic species such as protons feature a depth-dose curve that concentrates the dose around the Bragg peak, a characteristic of the Bethe-Bloch energy loss for these particles \cite{paganetti05,Linz:2012et,Peach:2012im,durantecharged2010,loefflercharged2013}.
The depth at which the peak occurs increases with particle energy; for incident protons above 70~MeV an approximate rule of thumb is that protons lose around 1~MeV per millimetre of water traversed, although this reduces with increasing incident energy.
The range of a 230~MeV proton is roughly 33 cm in water, so that this energy is sufficient to be used for tumour treatment in a typical adult patient.
For a given depth of the Bragg peak, heavier particle species such as carbon ions require greater energy; for example, to treat to a depth of 33 cm would require a C$\,^{6+}$ ion of 400~MeV/nucleon.
Both accelerators and beam delivery systems to the patient must cope with the significantly greater beam rigidity if carbon ions are used (see later), leading to greater magnetic bending and focusing fields and thus larger accelerators.

With efficient ion sources and compact accelerators, protons are the most accessible of hadronic beams and deliver tangible benefits over photons when traded-offs against accelerator size and cost \cite{paganetti05}.
However, heavier ions have the distinct advantages of greater radio-biological effectiveness (RBE) and smaller lateral scattering; carbon ions in particular have been used for patient treatment for this reason \cite{chu11}.

The sharp peak in the hadron depth-dose curve means that accurate imaging and planning are very important; a 1 cm range error in x-ray radiotherapy will change the dose to the tumour or normal tissue by approximately {2 \%} along the beam path, whereas a 1 cm range error in hadron therapy will shift the distal edge of the dose distribution and may thereby give a much larger change to the dose delivered at a particular location.

%a 2 mm range error can be accommodated in photon treatment planning, but the same range error for protons would risk serious damage to a sensitive structure lying close to the distal edge of the Bragg peak.

Whilst CT provides sufficient tissue density information for planning x-ray treatments, the conversion of Hounsfield numbers to tissue density is not accurate enough to provide optimal proton planning \cite{paganetti05}.
Hence there is significant interest in improving existing imaging techniques with, for example, a number of researchers developing proton computed tomography (PCT) \cite{hanson81,schneider94,schulte05,talamonti2010} and positron emission tomography (PET)-based dose monitoring, the latter utilising the $\,^{11}$C and $\,^{15}$O generated during proton or carbon irradiation \cite{parodi07}.

In order to provide PCT, particularly in adult patients, the proton energy must be larger than that required for treatment; protons exit the patient with a residual energy which is measured and - by comparison with the already-known entrance energy - used to determine the integral of the patient tissue density along the line between entrance and exit \cite{schulte04,schulte05}.
Tomographic reconstruction from many such tracks is undertaken in the same manner to other imaging techniques, but since the density determined here is proton-specific it may be translated directly into a required proton energy for treatment.
This method may thereby reduce range error if the proton energy can be measured accurately enough.

As well as photons, protons and carbon ions, radiotherapy has also utilised other particle species.
Very high-energy electron therapy (VHEET) uses an electron beam (of energy up to 250~MeV) directly rather than using it to create x-rays \cite{wideroe66,yeboah2002}; at sufficiently large energies there is believed to be a therapeutic benefit compared to photons, but compact facilities require larger accelerating field gradients than are yet commercially available.
Fast neutrons were used in clinical studies during the 1960s and 1970s but poor dose localisation overshadowed the high-LET benefits \cite{morgan72,catterall75,maor81,wambersie94}.
Finally, exotic species such as pions and antiprotons are also candidates for radiation therapy.
The potential advantage of these latter two species is the additional energy produced from the nuclear absorption of the stopped particles, adding to the energy released at the Bragg peak depth; this is referred to as a star dose \cite{li74}.

The use of antiprotons has also been studied, particularly at the Antiproton Cell Experiment (ACE) at CERN~\cite{Holzscheiter:2006eb}; the proposed advantages are the RBE enhancement adjacent to the Bragg peak from the antiproton annihilation, and the possibility of using the pions generated for dose monitoring.
Recent studies confirm such an RBE enhancement, but at present there is disagreement about the acceptability of the longer-range halo dose \cite{Kavanagh:2013br,Goitein:2008fh,Paganetti:2010cd}.

Held out as a great hope for treatment of glioblastoma, boron neutron-capture therapy (BNCT) is a combined technique in which epithermal neutrons (from a suitable reactor or accelerator-based source) are shaped in energy by a moderator assembly to optimise their spectrum to be absorbed by $\,^{10}$B nuclei bound chemically to (possibly dispersed) tumour cells, giving rise to alpha particle emission and hence a (very) local radiation dose \cite{taylor35}.
The main challenges for BNCT use are the availability of a compact, high-flux neutron source, the subsequent shaping of the source spectrum to avoid neutron irradiation that can cause very substantial damage to tissue outside the tumour region and the development of suitable boron compounds that can penetrate the blood-brain barrier and bind to tumour cells \cite{greenbnct98,barthbnct2005}.
Present BNCT facilities utilise either a neutron source derived via a graphite column from a nuclear reactor, or moderate the fast, broad-spectrum neutron output from a thick lithium or beryllium target struck by protons with energies typically between 2 and 30~MeV \cite{bluebnct2003}.

\section{History of Particle Therapy}	

Following Robert Wilson's suggestion of using protons for radiation therapy \cite{wilson47}, experimentation was started using the 184-inch synchrocyclotron at Lawrence Berkeley Laboratory \cite{lawrence57,tobias58}; the first patients were treated in 1952, but not with stopped beams \cite{JohnHLawrence:1962uh}.
At that time diagnostic devices were not available to provide adequate tissue-density information to determine the correct beam energy for the protons to stop in the tumour; this level of sophistication did not emerge until CT scanning became available in the 1970s.
The earliest treatments used plateau irradiation, in which the full energy beam of up to 900~MeV was collimated and passed through the target volume.
Incident beams were delivered to the patient at a variety of angles, overlapping the delivered fields to concentrate the dose at the desired location \cite{rajubook,castroepac}.
Target volumes were small (for example at the pituitary gland) and ablation studies for treatment of endocrine-related diseases (Cushings acromegaly) were highly successful.
Over 2000 patients were treated by this technique in Berkeley.
Russian programmes at ITEP and at St. Petersburg Nuclear Physics Institute/CRIRR were also quite successful \cite{wiesbook,Dzhelepov:1973vh,Chuvilo:1984ib,Abrosimov:2006de}.
Following experiments initiated at NIRS (Chiba) and at Tsukuba at the end of the 1970s \cite{Kanai:1980jw,Kurihara:1983ue}, Japan has become one of the world's largest users of both proton and heavy ion therapy \cite{Kitagawa:2010bj}.

Stopped-beam treatments with protons (i.e. utilising the concentrated energy loss at the Bragg peak depth) were started at the Gustav Werner Institute in Uppsala (Sweden) and at the 160~MeV Harvard Cyclotron (USA) in the early 1970s \cite{rajubook,bonnett93,chu93}, although stopped helium ions had been used a few years earlier \cite{JohnHLawrence:1962uh}.
Heavier ions were pioneered at the Berkeley accelerators, the 184-inch synchrocyclotron producing helium beams and the Bevalac generating therapeutic beams from carbon to argon.
Over 2000 patients were treated with these heavier ions, with the majority of Bevalac patients (close to 500) receiving treatments with neon ions.

The first dedicated hospital facility was developed at Loma Linda using a Fermilab-designed 250~MeV synchrotron and opened in 1990 \cite{lomalinda1,lomalinda2,Slater:1992gi,chu93}.
A synchrotron was chosen as it was considered that a synchrotron could provide the rapid energy variation and dose rate control required for accurate patient treatment.
Subsequently cyclotrons, synchrotrons and latterly other technologies were developed for proton therapy, initially at accelerator laboratories and then commercialised by a number of providers \cite{chu93,bonnett93}.
The Loma Linda facility pioneered the delivery of a single source beam to more than one treatment room, enabling multiple patient treatment, although not simultaneously due to the time required to switch from one room to another.
Loma Linda also demonstrated the clinical use of rotating gantry systems that deliver protons into a patient from a number of directions, which avoids the need to rotate the patient other than in the horizontal plane, whilst essentially still providing treatment beams from any angle.

Extensive trials with pion beams were conducted at Paul Scherrer Institute (PSI), TRIUMF and LAMPF in the 1970s, of which the most creative beam delivery concept was developed at PSI; called the Pion Concentrator (or Piotron) \cite{kaplan73}, the production target was struck with 590~MeV protons with 60 channels fanning out in a cone from the optimum production angle of about $60\,^{\circ}$.
These channels transported purified and energy-selected pions and brought the beamlets to an image point of the target at the site of the patient; a sufficient dose rate of pions could be achieved so that a daily fraction could be delivered in a few minutes.
Several thousand patients were treated with pions at these facilities \cite{fowler61,katz74,blattmann79,vonessen82}; clinical results were unsatisfactory, in part due to the relatively poor dose localisation from greater multiple scattering of the pions compared to the substantially heavier protons.
There have been studies of more cost-effective pion sources, such as PIGMI \cite{Knapp:1976tv}.

Heavy ions for therapy first became available in the early 1970s at the Princeton-Pennsylvania Accelerator \cite{Isaila:1970av,White:1971bz} and at the Bevalac in Berkeley \cite{Raju1978eh}, with measurements of the stopping profile and biological effects being made quickly \cite{Schimmerling:1971bb,Todd:1971hu}.
Being more massive than protons, heavy ions such as carbon, oxygen, nitrogen and neon will scatter less and thereby provide a sharper edge to the imparted radiation field \cite{Jackson1981ev}.
In addition LET is significantly higher than for protons.
%particularly in the Bragg peak and for carbon and heavier ions.
However, a significant proportion of the ions may undergo nuclear reactions that produce lighter fragments that can travel beyond the primary Bragg depth; this tail dose is significant when treating at depths greater than 20 cm.
%Long-term follow-up of more than 400 Bevalac patients showed that oxygen and neon LET were higher in the plateau region.
More recent facilities in Japan, Germany and Italy have opted for carbon ions, mainly because of the less damaging plateau dose and RBE enhancement near the Bragg peak, but also because of the reduced fragmentation and more manageable accelerator size.

%partly because of the Bragg peak RBE enhancement but also because of lower fragmentation and more manageable accelerator size.

\section{Particle Sources}

At present, nearly all hospital or laboratory centres use either a cyclotron or synchrotron as a particle source of either protons or carbon ions \cite{hiramoto07,flanz07,amaldi10}, delivering treatment beams to either single treatment rooms or (more frequently) multiple treatment rooms; in the latter case a transfer line and switching magnets are required to select which room receives beam.
Both cyclotrons and synchrotrons are mature technologies with well-established routes for delivering maximum proton energies in the 200-350~MeV range or even higher \cite{Papash:2012hb,Flanz:2013kt}.
At present, carbon ions at clinically useful energies are only provided using synchrotrons, although several cyclotron- and linac-based developments are underway.

Many alternative technologies are in active development; these include linacs, hybrids of linacs with circular machines \cite{amaldi07}, new techniques involving induction acceleration (e.g. dielectric wall accelerators), laser-generated proton beams \cite{linz07,Peach:2011ft}, or novel forms of conventional accelerators such as fixed-field alternating gradient (FFAG) rings \cite{peach10,craddock10,Peach:2013ft}.
Those technologies in or close to clinical implementation are described below, whilst some others of interest are discussed in Section 7.

\subsection{Normal-Conducting Cyclotrons and Synchrocyclotrons}

The normal-conducting cyclotron was the first circular (and therefore cyclic) accelerator to be used, following its invention by Ernest Lawrence in the 1930s \cite{Lawrence:1930wn,Lawrence:1931ff,Lawrence:1936dh}.
There are a number of variants, which include techniques to keep the driving radio frequency acceleration in synchronism with the orbit period \cite{Friesel:2009hn,Jongen:2010th,Jongen:963779}.
Synchronism is achieved either by modifying the pole shapes to make the orbits isochronous, or by varying (in the synchrocyclotron) the RF frequency during an acceleration pulse.
Virtually all hospital-based therapy cyclotrons utilise the first approach and at present deliver a maximum proton energy of 235~MeV, but several commercial vendors have proposed or are constructing synchrocyclotrons.
Higher energies have been achieved in research machines such as the 590~MeV PSI cyclotron \cite{Wagner:2009jo}, but cyclotrons of this energy have not yet been commissioned outside of accelerator laboratories.

To allow for depth scanning of the tumour, some method must be employed to vary the proton beam distal (maximum) depth, since the output energy from a cyclotron is to all intents fixed.
To achieve this variation in incident proton energy, the beam from the cyclotron is passed through an adjustable thickness of some material, typically two back-to-back wedges of carbon.
The mean energy loss is accompanied by an increase in energy spread and beam emittance due to scattering, so the output beam must be cleaned in an energy-selection system prior to the final beam delivery system.
Energy selection is achieved with a combination of dipole magnets, collimators and extensive shielding to select the energy and emittance of the beam that can be transported through the gantry to the patient.
Typically cyclotrons can deliver sufficient beam intensity so that despite the beam loss from the energy selection process (by as much as a factor of one hundred or more for the lowest energies), they provide a dose rate that is competitive with other accelerator types.
It is also worth noting that in typical cyclotron designs the output spot is basically circular, which simplifies the beam coupling to the final beam delivery system, for example to a downstream rotating gantry.

\subsection{Superconducting Cyclotrons and Synchrocyclotrons}

The higher fields (above about 3~T) which are available from superconducting magnets allow higher extraction energies at a particular orbit radius \cite{Blosser:1989ve,Schillo:2001ut,Schippers:2004wf,Roecken:2010vs,Jongen:2010eb}; this allows cyclotrons to be made smaller for a given extraction energy but with the penalty of the greater capital cost of the magnet and operating cost of the cryosystem \cite{Papash:2012hb,Alonso:2012bi}.
Reliability is not necessarily worse with a low-temperature rather than a room-temperature magnet, as in practice reliability is often determined by other points of failure.
However, maintenance work on the accelerator will usually include a significant time overhead from warm-up and cool-down of cryogenically-cooled components.

Several commercial vendors are now either developing or offering superconducting cyclotrons \cite{Jongen:2010eb}; Mevion utilise NbSn$_3$ superconducting magnet coils that allow the cyclotron to be small enough to be gantry mounted whilst maintaining a maximum proton energy of around 250~MeV.
Varian-ACCEL use NbTi to have a larger, lower-field and therefore simpler magnet to obtain 250~MeV protons at extraction \cite{kimvarian}, whilst IBA also use NbTi in their 230-250~MeV S2C2 cyclotron \cite{Alonso:2012bi}.
Superconducting magnets are often less than half the weight of their normal-conducting equivalent \cite{Linz:2012et}.
There is a research and commercial push to develop cyclotrons with higher extraction energies, but as yet no commercial solution has been installed to our knowledge.

\subsection{Radiofrequency Linacs and Cyclinacs}

Proton linacs for radiotherapy have been proposed, using either a conventional pre-injector or using a low-energy cyclotron as the proton source (so-called cyclinacs).
Accelerating structure developments have been made towards the goal of having sufficient accelerating gradient for them to be used in a hospital context, particularly at S-band (3 GHz) frequency; two Italian collaborations - TERA \cite{Amaldi:2004ul,Amaldi:2009wa,DeMartinis:2012fr} and TOP-IMPLART \cite{Ronsivalle:2011za} - have separately developed 3 GHz structures \cite{Amaldi:2009wn}, whilst other groups have studied upgrades of existing cyclotrons \cite{ref20}.
The frequency mismatch between the cyclotron and linac can result in significant beam loss in the first cells of the linac, but the higher current available at low energy compensates for this.
This technique has also been pursued for carbon-ion therapy \cite{Garonna:2010ws} with structure development underway \cite{Degiovanni:2011fq,VerduAndres:2013iy}.
Low-energy sections for proposed facilities have been constructed in Italy and clinical facilities are either planned or proposed here and elsewhere.
The TERA TULIP design proposes combining the accelerating structures and rotating gantry to provide a single-room solution \cite{Degiovanni:2013vi}.
The same structure development has also been recently proposed by Advanced Oncotherapy to provide a linac-based proton therapy centre \cite{avoadamrelease}, although no centres are yet operating using this scheme.

\subsection{Synchrotrons}

Synchrotrons are a well-established technology in proton therapy and there have been numerous slow-cycling (up to a few hertz) examples used clinically \cite{lomalinda1,lomalinda2,chu93,Badano:385378,Badano:1999ew,Bryant:1999hy,Umegaki:2003va,Amaldi:2004it,Haberer:2004im,Tuan:2013td}.
At present, synchrotrons are the only accelerator technology used to provide carbon ion beams.
All current designs follow the original approach by GSI and implemented at Heidelberg \cite{Haberer:2004im}, which uses: an ECR ion source, a radio-frequency quadrupole and drift-tube linac, and finally a synchrotron to accelerate carbon ions up to 400~MeV per nucleon \cite{Hirao:1992kq,Sato:1995td,Noda:2007ee}.
Energy variation of the proton or carbon beam distal (maximum) energy is accomplished by extracting the beam at different times in the acceleration cycle.
Extraction is commonly achieved using either a third-order resonance or RF-knockout scheme and a feedback system is employed to ensure that the extracted dose can be controlled to provide both good dose stability and beam intensity that is programmable in time.
Furthermore, extraction can be gated to synchronise the dose with the patient's breathing, in order to minimise the effect of organ motion.
In comparison with isochronous cyclotrons, where particles are continuously injected and accelerated, in synchrotrons (and similarly in synchrocyclotrons) the injection window is short, so the number of particles available at treatment energies is substantially lower.
Nonetheless, dose rates are still adequate and in a synchrotron the available particle flux at each extraction energy is essentially constant; this is in contrast to cyclotrons, where the the extracted beam intensity is reduced by up to three orders of magnitude by the scatterers used to reduce the mean particle energy.

\subsection{Rapid-Cycling Synchrotrons}

Brookhaven National Laboratory (BNL) in conjunction with a commercial company (BEST) \cite{Trbojevic:2011tk} have spearheaded the idea of a rapid-cycling synchrotron for either proton or carbon-ion therapy, although other groups have also initiated designs \cite{AlHarbi:2003ba}.
BNL propose that a maximum cycling frequency of 30 Hz allows the use of simpler magnets and resonant power supplies and in principle a pulse-by-pulse energy variation at the cycling frequency is possible by timing the firing of the extraction kicker during the acceleration phase \cite{Peggs:2002vq,Cardona:2003tq}.
Slow extraction is not possible and intensity feedback is both required and difficult; intensity control must be done either by trimming during acceleration, or by exquisitely fine control over the injected current.

\section{Facility Layout Options}	

A single accelerator can provide a particle beam to a single treatment room or to multiple treatment rooms; some rooms may then have gantries to translate the horizontal beam entering the treatment room into the vertical plane, with the patient either stationary with the gantry rotating around them (isocentric) or with some combination of patient and gantry motion.
The rationale for gantries, and their design issues, are discussed below.
The combination of a rotating gantry with a patient table that rotates in the horizontal plane allows the particle beam to enter the patient from any angle, in contrast to treatment with a fixed beam orientation.
A third option is to mount the accelerator on the gantry, which is only possible if the accelerator is small and light enough that the resulting gantry is of reasonable size; of course in this scenario the accelerator can only support one treatment room.

%A notable counter-example is the Mevion gantry-mounted cyclotron, of which several examples are presently being installed in the USA.
In most proton and carbon-ion systems today the particle source is not mounted on the gantry, but instead is located in a separate, static location.
The advantage of having several treatment rooms supplied by a single accelerator is that patients can be prepared for treatment in parallel prior to irradiation, thus making optimal use of the available accelerator time.
The gantry beam-optics system must couple to the source, and beam switching is ideally fast so that the beam is ready for initiation of treatment within a short time (typically less than one minute) from the operator's request.
The disadvantages of a single accelerator/multi-room arrangement are the reduction in flexibility and the complication of parallel patient scheduling, although the optimisation of throughput planning is becoming more advanced.
Monte Carlo modelling studies suggest three to four treatment rooms per accelerator provides optimal utilisation of all resources \cite{Goitein:2003dj,Fava:2012ip,Aitkenhead:2012gx}.
In a multi-room, single-accelerator facility there is the risk of the accelerator source being the single point of failure for all treatment rooms, which must be taken into account.
Reliability (measured as the fraction of time available over that demanded) of the accelerator systems of at least 98 \% is typically required \cite{Ma:2012um}.
Such facilities may be upgraded by modifying or replacing the accelerator whilst leaving the treatment rooms as they were; this modification has been performed at the PSI proton therapy centre where a dedicated cyclotron was added to an existing treatment room suite \cite{Schippers:2006hy}.
%, but not performed at a hospital-based treatment centre.

Whilst the shielding requirements in all proton or carbon-ion facilities are naturally significant, particular attention must be paid in cyclotron-based facilities to the relatively larger amount of activation arising from the energy-selection system.
Conversely, coupling cyclotron and linac beam optics to gantry optics is relatively simpler due to the circular beam spot.
This is in contrast to synchrotrons where the spot may be significantly asymmetrical; this requires rather special measures to match the accelerator and gantry beam optics \cite{Benedikt:783137}.

If the particle source can be made small enough the option exists to mount it directly on the treatment gantry.
Since the accelerator is directly mounted on the gantry, the beam optics are much simpler but the gantry as a whole may not be smaller or lighter, depending on the source used (typically a compact cyclotron \cite{bloch11, bloch12}, but potentially a dielectric wall accelerator or laser-based acceleration scheme, see below).
The issues of parallel treatment and scheduling are greatly simplified or eliminated and if there are a number of treatment rooms each with its own gantry and source then there is no beam-derived single point of failure.
Beam clean-up and elimination of the parasitic neutron dose will however be more complex as they must be done close to the patient.
Whilst some studies indicate that this could be manageable, the issue is still a the subject of research \cite{MeasuredNeutronLev:2013vb}.
Presently, only compact superconducting synchrocyclotrons can deliver 250~MeV protons whilst at the same time being small enough for the gantry and treatment room to be practicable.

\section{Beam Delivery and Field Formation}

%\subsection{Introduction}

It is of course not enough to merely bring a hadron beam into a treatment room; its distribution transversely and longitudinally must be capable of being conformed to the tumour volume in $(x,y,z)$ according to a predetermined treatment plan.
Prior to the operation of the Loma Linda facility, treatments were carried out using both fixed beamlines and fixed patient orientation.

Flexibility in the entry orientation of the beam is important, and whilst static horizontal delivery is still preferred for some fields (such as ocular treatments) the capacity to provide an arbitrary entry angle, via a gantry, is highly favoured by the medical community \cite{Blattmann:1992ex}.
The use of gantries and the related issues of field shaping and dose conformation and how they are achieved using passive or active methods are discussed below.

\subsection{The Use of Gantries}

As well as the requirement to spread the dose throughout the volume of the tumour, it is of course desirable to direct treatment beams with the least possible impact on organs or other tissues that are either in close proximity to the treatment volume or which lie between the treatment volume and the patient's external surface.
To achieve this goal, it is necessary to either rotate the patient or rotate the beam line that is delivering the particles to the patient.
Although it is acceptable to rotate or translate a supine patient in the horizontal plane, full six-dimensional patient motion is not possible \cite{Blattmann:1992ex}.
This is primarily due to the undesirable organ motion relative to the external body reference points used to align the treatment beam; thus horizontal rotation is acceptable but not other axes.
Rotation may also be difficult for the patient, or interfere with other medical procedures such as anaesthesia.
Some treatments are suitable for use with fixed angle beam (which may be horizontal, vertical or something in between), with the patient either supine, prone or sitting up; a notable example is the treatment of ocular tumours where the patient is seated and the incident particle beam fixed and horizontal.
However, to provide maximum flexibility the concept of the rotating delivery beam line (gantry) has been developed, mimicking the situation in x-ray radiotherapy where the gantry rotates $360 \,^{\circ}$ around a patient lying on a couch, with the couch's default position along the axis of gantry rotation.
If the patient couch is allowed to rotate in the horizontal plane, a more compact treatment room is possible as the gantry is only required to rotate $180 \,^{\circ}$ (from vertically-down to vertically-up).

There are several reasons to keep the patient supine during radiotherapy treatment:
\begin{itemize}
  \item To more easily reproduce the patient position used for pre-treatment imaging such as CT that has been used for treatment planning;
  \item To more easily reproduce the patient position through several weeks of daily treatment;
  \item To maximise the flexibility in possible beam directions to avoid normal tissue and therefore to target the tumour most effectively.
\end{itemize}

In external beam radiotherapy by far the most common method is for the patient to lie on a treatment couch, imitating the position used for CT scanning.
When accompanied with sensible immobilisation devices, this ensures organs are in similar position during both treatment planning and treatment, to give a reproducible set up.
To maximise the number of possible treatment angles the couch can have up to five degrees of freedom, with tilts of the couch limited to less than $5 \,^{\circ}$ for patient comfort.

%the use of a variable field angle may be used to avoid shadows from bones or implants between the tumour and the patient surface [ref needed].

\subsection{Gantry Design}

\begin{figure}[htb]
\centerline{\includegraphics[width=0.8\textwidth]{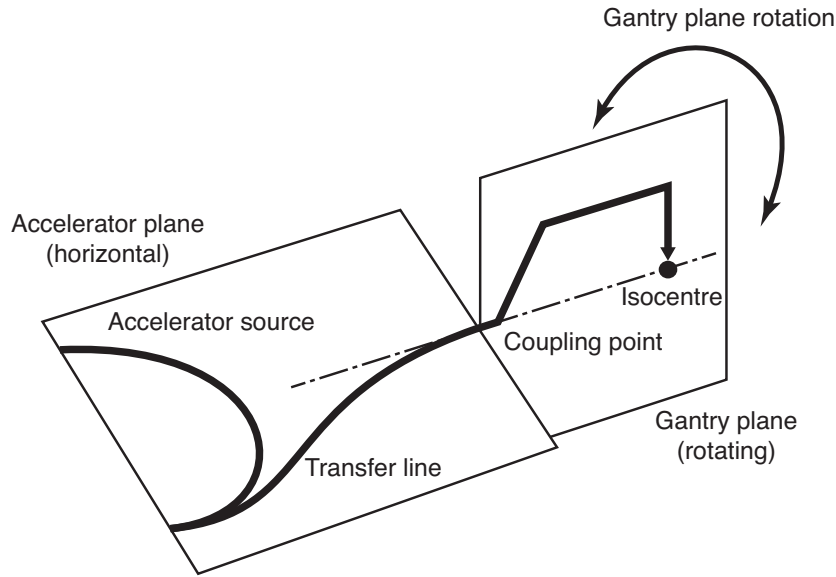}}
\caption{A schematic illustration of a particle therapy installation incorporating an isocentric particle therapy gantry. The gantry plane rotates around an axis that is horizontal and passes through the isocentre within the patient. \label{isocentric}}
\end{figure}

Conventionally, the term gantry is used to refer to a particle beam transport line that is designed to rotate around an axis; a gantry consists of a mechanical support structure, drive mechanism, magnets, vacuum vessels and beam diagnostics, along with other secondary technical infrastructure.
The de-facto standard gantry design that has emerged is the isocentric approach, wherein the patient is kept essentially still in a prone position with the centre of rotation of the gantry passing in the horizontal plane through the patient and the incident beam rotated in the vertical plane around the patient.
A number of dipoles are used to transport the beam from the particle source to the treatment room which it enters along this axis.
Further dipoles on the gantry are used to create an offset between the beam and the gantry rotation axis and finally turn the beam toward the patient who is positioned at the isocentre.
Quadrupoles are used to provide beam focusing both in the beam transport prior to, and within, the gantry.
The isocentric approach is shown schematically in Fig. ~\ref{isocentric}.

The components between the final bending magnet and the exit window to the patient are sometimes referred to as the treatment nozzle.
As well as beam position and dose monitors it may include deflection magnets to scan the spot transversely, or alternatively beam scatterers and collimators to spread and shape the proton dose across a chosen field size; these field sizes may now be as large as 30 $\times$ 40~cm.
The need for a treatment nozzle will always require a distance of several metres between the last bending magnet of a gantry and the patient.
To obtain a full $4 \pi$ coverage of the patient, the gantry must rotate at least $180 \,^{\circ}$ in conjunction with a patient table that rotates $360 \,^{\circ}$ in the horizontal plane.
To minimise patient rotation (which introduces time delay and may result in unacceptably large patient misalignment) many gantry designs utilise the full $360 \,^{\circ}$ of rotation around the patient.

\begin{figure}[htb]
\centerline{\includegraphics[width=0.8\textwidth]{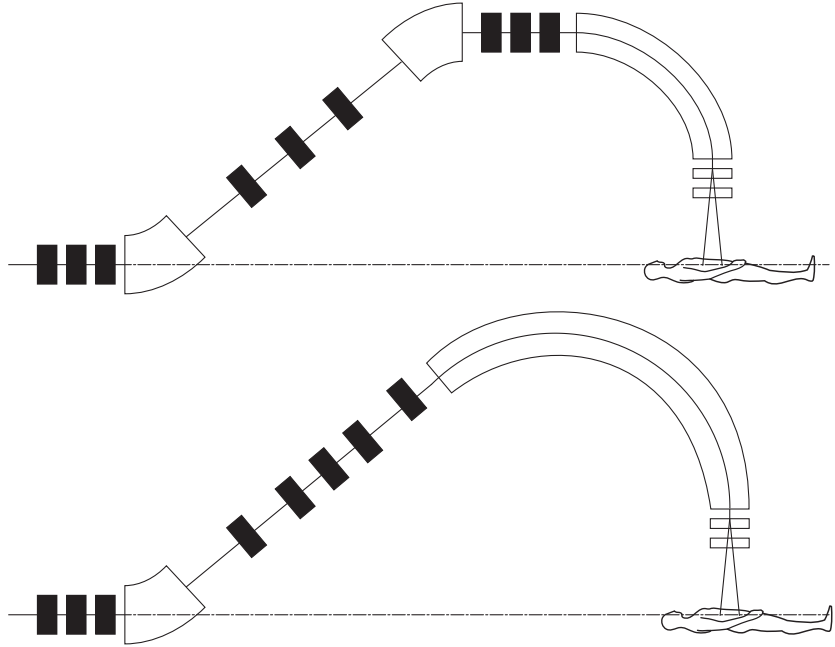}}
\caption{Schematic illustration of isocentric 2- and 3-dipole gantry designs. The number of quadrupoles required may be different to that shown. In both cases shown, spot scanning at the patient is obtained by using fast scanning magnets placed downstream of the final dipole. As the energy of the protons is varied the gantry magnetic field strengths must be scaled to match the proton beam rigidity. \label{TwoThreeDipole}}
\end{figure}

Isocentric gantry designs typically incorporate either two or three dipoles as illustrated in Fig.~\ref{TwoThreeDipole} (but may incorporate more \cite{Noda:2007ee,Furukawa:2008ef}) and are designed to minimise the total mass of the gantry magnets whilst also minimising the outer radius and length of the entire rotating structure.
Despite this, gantries still involve a large mass of magnet steel and a mechanical structure that is many metres in both length and radius that rotates around the patient; most of this assembly is typically disguised behind a false wall and therefore not visible to the patient in the treatment room.
Gantry-mounted sources such as compact cyclotrons do not require the same number of beam steering elements, since the particles from the cyclotron may be directed straight at the patient; however there is then less space to incorporate beam cleaning and scanning elements.

\begin{figure}[htb]
\centerline{\includegraphics[width=0.8\textwidth]{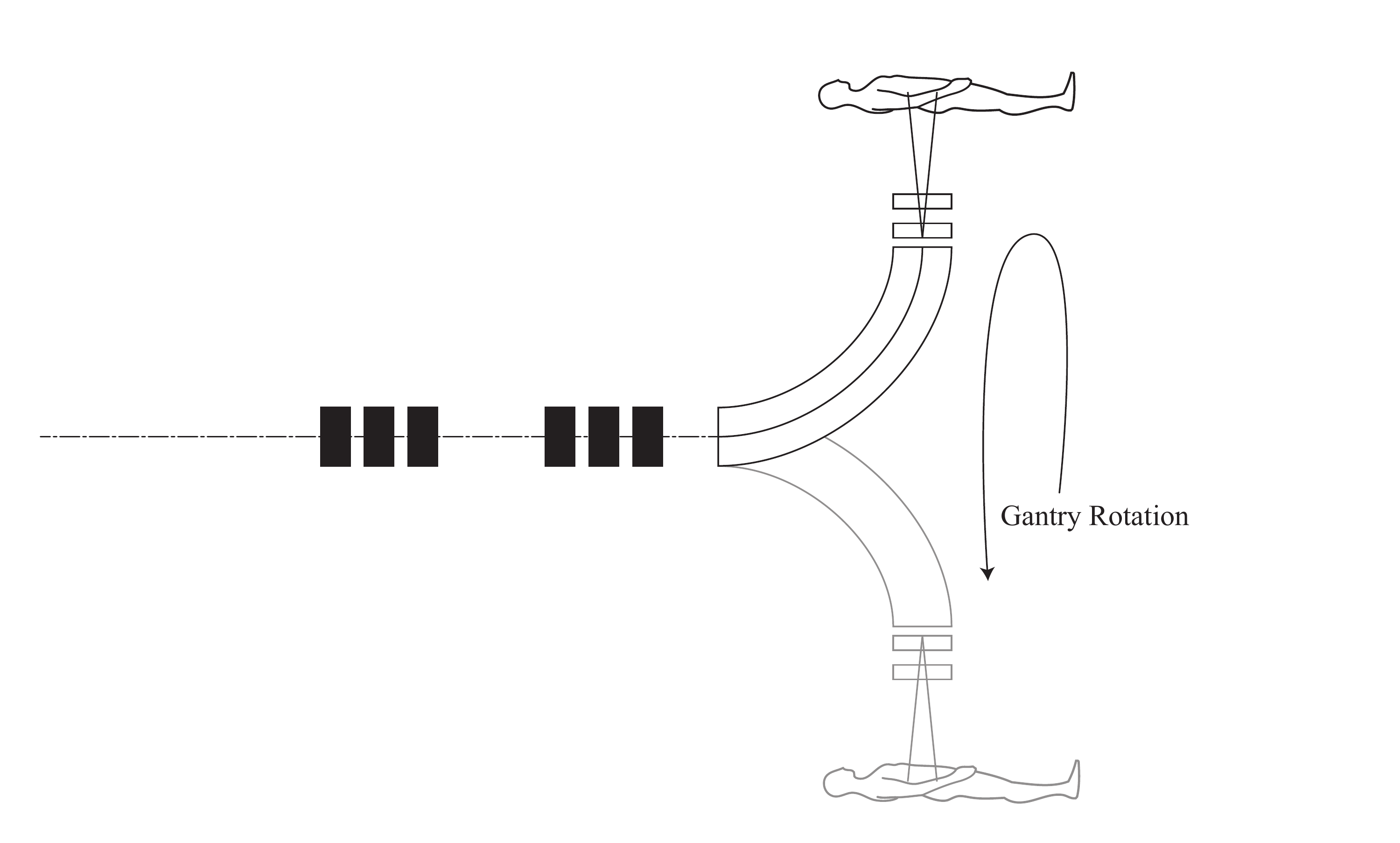}}
\caption{Schematic illustration of Riesenrad gantry.
\label{Riesenrad}}
\end{figure}

Alternative approaches to the isocentric concept have been proposed to reduce the mass of moving steel.
The most notable of these is the so-called ``Riesenrad" (or ferris wheel) gantry, wherein a smaller number of deflecting magnets lie on the gantry rotation axis and the patient displaced from that axis; if the gantry and patient are rotated (but the patient kept supine) the patient will receive the dose from different angles \cite{Benedikt:1999ip,Reimoser:2001ir}.
The Riesenrad gantry is shown schematically in Fig.~\ref{Riesenrad}.
Such gantries have less moving steel and are therefore lighter, but have more complex patient-handling issues compared to isocentric gantry installations; the principal of these is that patient entry and egress are more time-consuming, which may be an issue during emergencies.

A hybrid scheme (often referred to as an eccentric gantry) has been implemented at PSI (at Gantry 1), in which both the patient and the beam line magnets move around the rotation axis, thus minimising the overall diameter of the gantry \cite{pedronigantry1,pedronigantry2}.
Unlike the Riesenrad gantry, which requires a larger shielded room, the compact eccentric gantry design reduces the swept volume to a minimum.

Most gantries in commercial proton therapy systems are isocentric and their size is dictated by their use of normal-conducting dipole magnets, whose maximum field on the beam axis is restricted to around $\approx$1.8~T.
Given that the beam rigidity of protons at 250~MeV is 2.43~Tm, a 1.8~T field implies a bending radius of 1.35~m; this bending radius, along with the space needed for the beam-spreading system to cover the treatment field, sets the overall size of proton gantries, which are typically 5 to 6 metres in radius and have masses between 100 and 200 tons.

\begin{figure}[htb]
\centerline{\includegraphics[width=0.8\textwidth]{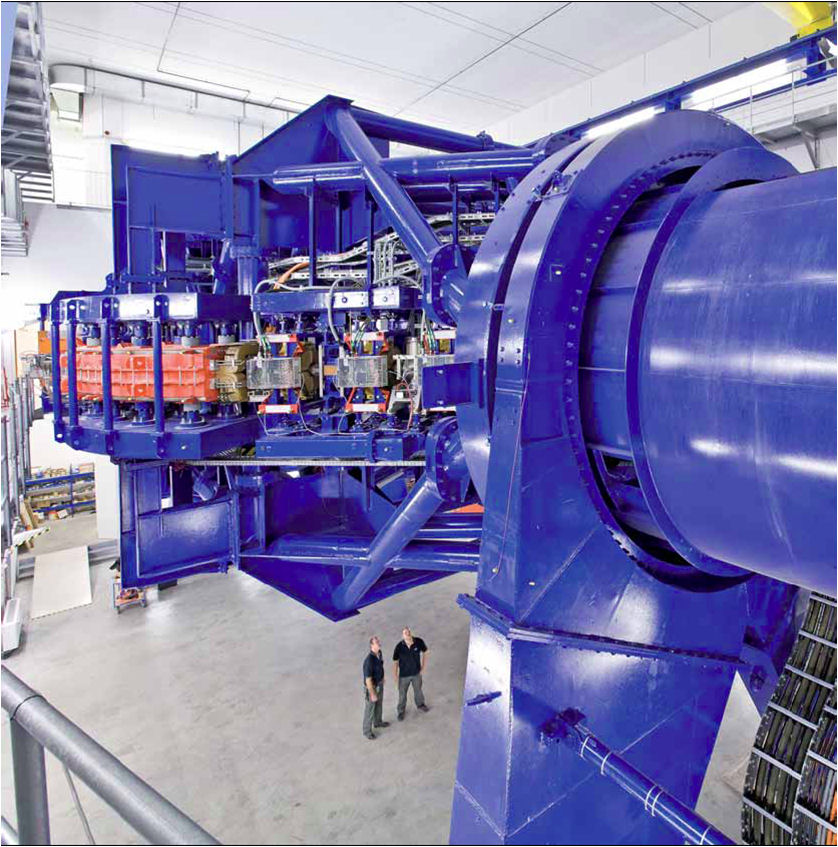}}
\caption{The Heidelberg carbon-ion gantry, presently the only operating gantry for carbon ion treatment. (courtesy Prof. Thomas Haberer/Heidelberg Ion Beam Therapy Centre)\label{hitgantry}}
\end{figure}

For carbon ions the gantry must be significantly larger.
The only current example of a carbon gantry is situated at the Heidelberg Ion Beam Therapy Centre (HIT), and is shown in Fig.~\ref{hitgantry} \cite{Haberer:2004im}.
Here the maximum carbon ion energy used for treatment is 425~MeV/u, corresponding to a beam rigidity of 6.57~Tm and a bending radius of 3.65~m.
The use of normal-conducting dipole technology produces a gantry $\approx$19~m in length and nearly $\approx$15~m in diameter and with an overall mass of 600 tonnes (of which 135 tonnes is in the magnets).
Despite the very large rotating mass, the positional accuracy of the isocentre is maintained to $\approx$1~mm, similar to that achieved in a proton gantry.

Another important factor in the gantry design is the method chosen to spread the energy of the delivered particles to cover the depth range of the treatment volume.
The beam energy can be chosen to position the Bragg peak at the distal edge of the tumour; range modulators in the treatment nozzle are then employed to create a spread-out Bragg peak (SOBP) to match the tumour depth profile.
Alternatively, if the dose is being delivered in a series of depth slices the energy of the particles from the source is varied.
In this latter case, the gantry magnet fields could in principle be kept fixed, as long as there is sufficient aperture to allow the required range of energies to be transported (for example in an FFAG gantries with a large energy acceptance, discussed later), but typically in practice the gantry magnet field strengths are varied to match the changing beam rigidity.
The use of variable gantry magnet fields does not usually limit the rate at which treatment may progress from one depth layer to another in the treatment volume, as it is usually faster than the rate at which the energy may be varied at the source (e.g. the energy selection system of a cyclotron); typically a range step of 5~mm may be achieved in $\approx$100 ms using a degrader, e.g. a pair of carbon wedges moving against each to give a varying thickness that is uniform across the incident proton beam.

\subsection{Scattering Methods}

The mechanism employed to distribute the dose throughout the tumour has a significant impact on the gantry design.
Coverage of the tumour by the beam in most of today's installations is achieved using a passive beam-spreading technique.
Passive scattering typically uses two scattering layers, in which the first (primary) scatterer spreads the beam out laterally and a second, about half the distance between first scatterer and patient, has a complex variation of thickness with distance from beam axis to provide a dose that is uniform in intensity across the treatment field width \cite{Koehler:1977du,chu93}.

With a double-scattering design, lateral conformation to the tumour can be achieved using either multi-leaf collimation (MLC), analogous to that used in x-ray radiotherapy systems, or by manufacturing a custom collimator for each individual patient treatment.
However, the use of such collimation in the treatment nozzle can produce an additional neutron dose for the patient.
An alternative approach to scattering is to use a set of so-called wobbler magnets that paint the beam in a series of concentric circles, along with collimators to establish the lateral field shape~\cite{chu93,FlanzPAC05}.

\begin{figure}[htb]
\centerline{\includegraphics[width=0.4\textwidth]{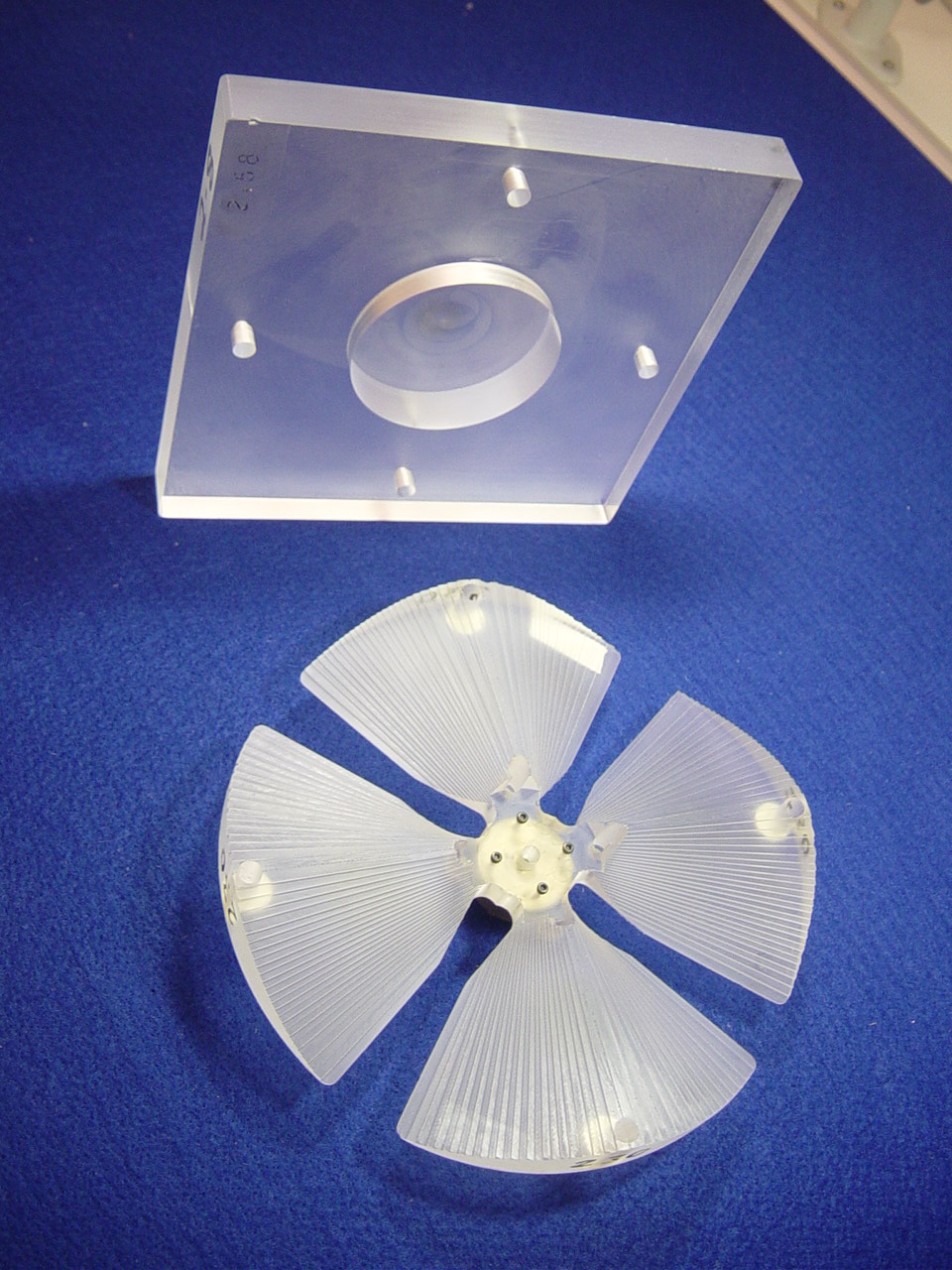}}
\caption{Compensator (top) and range modulator (bottom) used at Clatterbridge Hospital to provide longitudinal conformation of the 62~MeV protons delivered from their cyclotron to the depth range of the tumour. The modulator wheel is 190~mm in diameter. (courtesy Prof. Andrzej Kacperek/Clatterbridge Hospital)\label{ccomodcomp}}
\end{figure}

Longitudinal conformation is achieved using the well-established technique of utilising a range modulator and compensator \cite{Koehler:1975be}, the latter being manufactured specifically for each patient treatment.
These are typically manufactured from a polymer, the modulator being a spinning, wedged wheel, generating a SOBP covering the tumour longitudinally; an example of this is shown in Fig.~\ref{ccomodcomp}.
The compensator is a fixed-energy degrader conformed in thickness laterally across the tumour to have the distal edge of the SOBP follow the distal edge of the tumour.
This arrangement does not provide optimal conformation to more complex target volumes, since the constant depth range delivered may be larger than the tumour in some locations.
To achieve three-dimensional conformation, a range-stacking procedure must be used; the treatment volume is divided into a series of depth slices, each with a given outline determined by the MLC settings.
By treating each depth slice sequentially, an irregular treatment volume can be covered with higher precision.

\subsection{Spot Scanning}

The alternative to the scattering or wobbler magnet approaches is spot scanning.
This is a key emerging technology that is being developed for future proton and carbon therapy systems, as well as for retrofitting to existing facilities.
In most designs a pencil beam of a given energy and a few millimetres transverse size is directed to a given set of coordinates in the treatment volume by a pair of fast-scanning deflection magnets, which cause the beam spot to dwell on that voxel until the prescribed dose is delivered there.
The scanning magnets then move the beam to the next set of coordinates and so forth until the beam has been painted over the full area of the treatment volume at that desired depth.
The beam energy is then changed to treat the next layer of the treatment volume, gradually building up the layers until the entire volume is treated.
A treatment plan based on patient imaging is used to optimise the dose over the entire treatment volume, taking account of the dose both upstream and downstream of the Bragg peak delivered by a particular spot.
Depending upon the largest intended treatment volume the scanning beam may need to address transverse fields up to $400 \times 400$~mm, although most facilities presently specify not more than $200 \times 200$~mm.
In any case the treatment of very wide areas may be achieved by using the technique of field patching, in which the patient table is moved relative to the isocentre.

\begin{figure}[htb]
\centerline{\includegraphics[width=0.8\textwidth]{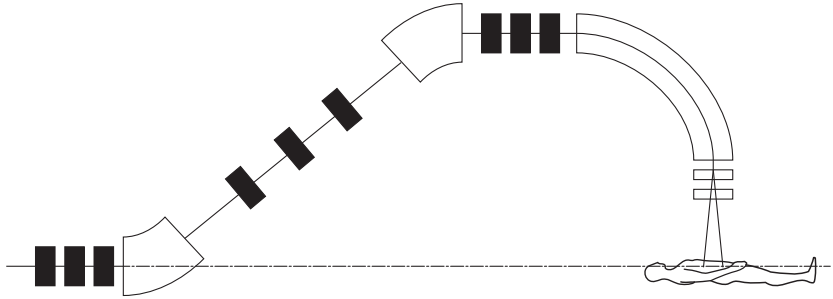}}
\caption{A schematic illustration of an idealised isocentric 3-dipole gantry which incorporates spot scanning magnets \textbf{downstream} of the final dipole. Whilst the aperture in the final dipole may be made relatively small, there may be a significant difference in field size (and correspondingly higher dose per unit area) at the patient surface compared to that at the isocentre due to the finite source-to-axis distance (SAD). \label{downstream}}
\end{figure}

The combination of pencil-beam spot scanning with a variable beam angle from a gantry enables the technique of intensity-modulated particle therapy (IMPT), in which several treatment fields (gantry angles) with inhomogeneous doses are combined to maximise the dose at the tumour and to minimise the dose to healthy tissue.
IMPT involves active beam delivery devices and thus requires similarly fast and accurate dosimetry; typically only a few milliseconds can be budgeted to deliver the dose to a voxel if the total treatment time is to be kept to a few minutes.
Other advantages of spot scanning are the elimination of patient-specific hardware - rendering patient setup faster and less costly - and a reduction in beam loss and hence neutron generation in the vicinity of the patient.
In most current treatment nozzle designs, the pair of fast-scanning deflection magnets are mounted after the final dipole, as shown in Fig.~\ref{downstream}.

An important figure of merit in pencil-beam spot scanning is the source-to-axis distance (SAD), which is the distance from the isocentre to the apparent source of the particle beam
If there are no beam focusing elements between the scanning magnets and the isocentre, the effective source location is the point at which changes to the angle of the beam are made, i.e. the location of the scanning magnets.
Since the beam trajectories diverge from the source, the area intercepted at the patient's skin will be smaller than the area at the isocentre depth, and thus a higher dose will be delivered to the skin.
Making the SAD as large as possible will therefore give the best degree of skin sparing.
There are two methods by which the SAD may be made either very large, or even infinite.
The first is to have two pairs of fast-scanning deflection magnets mounted after the final dipole (one pair for each plane) which could produce a parallel beam translation at the patient surface; this either requires a large distance from the final dipole to the patient to develop the lateral spot offset - which demands a larger gantry radius - or very technically demanding scanning magnets.
Whilst this double-magnet approach has been proposed - for example in combination with an FFAG gantry \cite{trbojevic11} - it has not yet been implemented anywhere.

\begin{figure}[hbt]
\centerline{\includegraphics[width=0.8\textwidth]{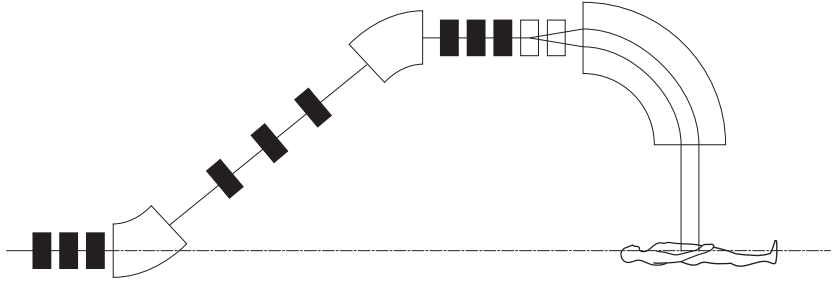}}
\caption{A schematic illustration of an isocentric three dipole gantry which incorporates spot scanning magnets \textbf{upstream} of the final ($90 \,^{\circ}$) dipole. With appropriate beam optics and entrance and exit edge angles, the betatron phase advance between the scanning magnets and the isocentre may be made approximately $90 \,^{\circ}$, resulting in an effectively infinite SAD. The disadvantage is that the aperture within the final dipole must be sufficient for the intended field size, necessitating a larger good field region and hence greater mass and excitation current requirements. \label{upstream}}
\end{figure}

\begin{figure}[hbt]
\centerline{\includegraphics[width=0.8\textwidth]{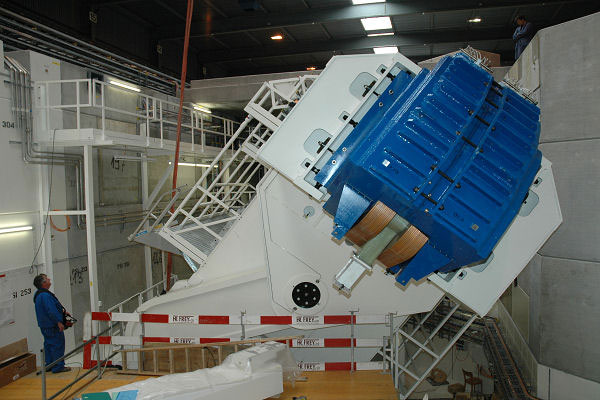}}
\caption{The gantry mechanism and final dipole of Gantry 2 at the Paul Scherrer Institute. The upstream design enables parallel scanning, which thereby simplifies treatment planning and gives an infinite SAD which assists in skin sparing but necessitates a rather large 45-tonne final dipole to give sufficient aperture to deliver the desired treatment field size. (courtesy Prof. Tony Lomax/PSI)\label{psigantry}}
\end{figure}

\begin{figure}[hbt]
\centerline{\includegraphics[width=0.8\textwidth]{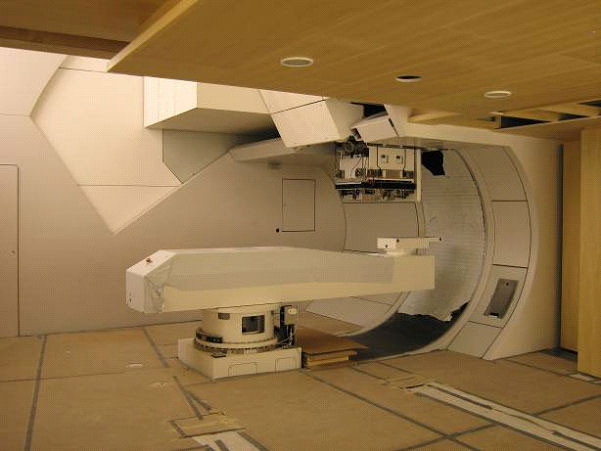}}
\caption{The Gantry 2 treatment room at the Paul Scherrer Institute. The rotating gantry is concealed behind a sliding false wall; sufficient distance must be provided between the end of the treatment nozzle and the patient to allow safe rotation. However, the distance from the end of the nozzle to the patient surface should also be minimised as far as practicable to limit scattering in the intervening air. (courtesy Prof. Tony Lomax/PSI)\label{psitreatment}}
\end{figure}

The second approach is to have either one or both of the fast-scanning deflection magnets upstream of the final dipole (one for each plane); this scheme is shown schematically in Fig.~\ref{upstream}.
The beam optics can then be designed such that the beam leaving the final dipole is translated parallel to the zero-deflection axis as the scanning magnets alter the entrance angle of the beam into the dipole, thereby giving an effectively infinite SAD.
In this upstream scheme the gantry radius does not need to be increased to include scanning; however the good field region (and therefore the magnet pole width and the pole gap) of the final dipole must be increased to allow for the transverse beam displacement required to produce the chosen treatment field size, which adds significantly to its weight.
In principle other beam optical arrangements may also be used to achieve the point-to-parallel focusing.
The first example of a parallel scanning gantry was PSI Gantry 2 \cite{pedroni04}, implemented in 1996; the final dipole is shown in Fig.~\ref{psigantry}, whilst the treatment room is shown in Fig.~\ref{psitreatment}.

The first proton treatment gantry was constructed at the Loma Linda proton treatment centre and utilised a now-unusual ``corkscrew" optics scheme, the design providing a full $360 \,^{\circ}$ rotation of the beam axis around the patient \cite{koehler1987,schulze91}.
Most commercial gantry designs presently offer $360 \,^{\circ}$ coverage and, if present, place the scanning magnets after the last dipole.
As scanning systems are usually retrofits to gantries originally fitted with passive scattering nozzles, there is an inherently finite SAD.
However, retrofitting a scanning nozzle downstream of the final dipole does not necessitate significant changes to the gantry beam optics.

\begin{figure}[htb]
\centerline{\includegraphics[width=0.8\textwidth]{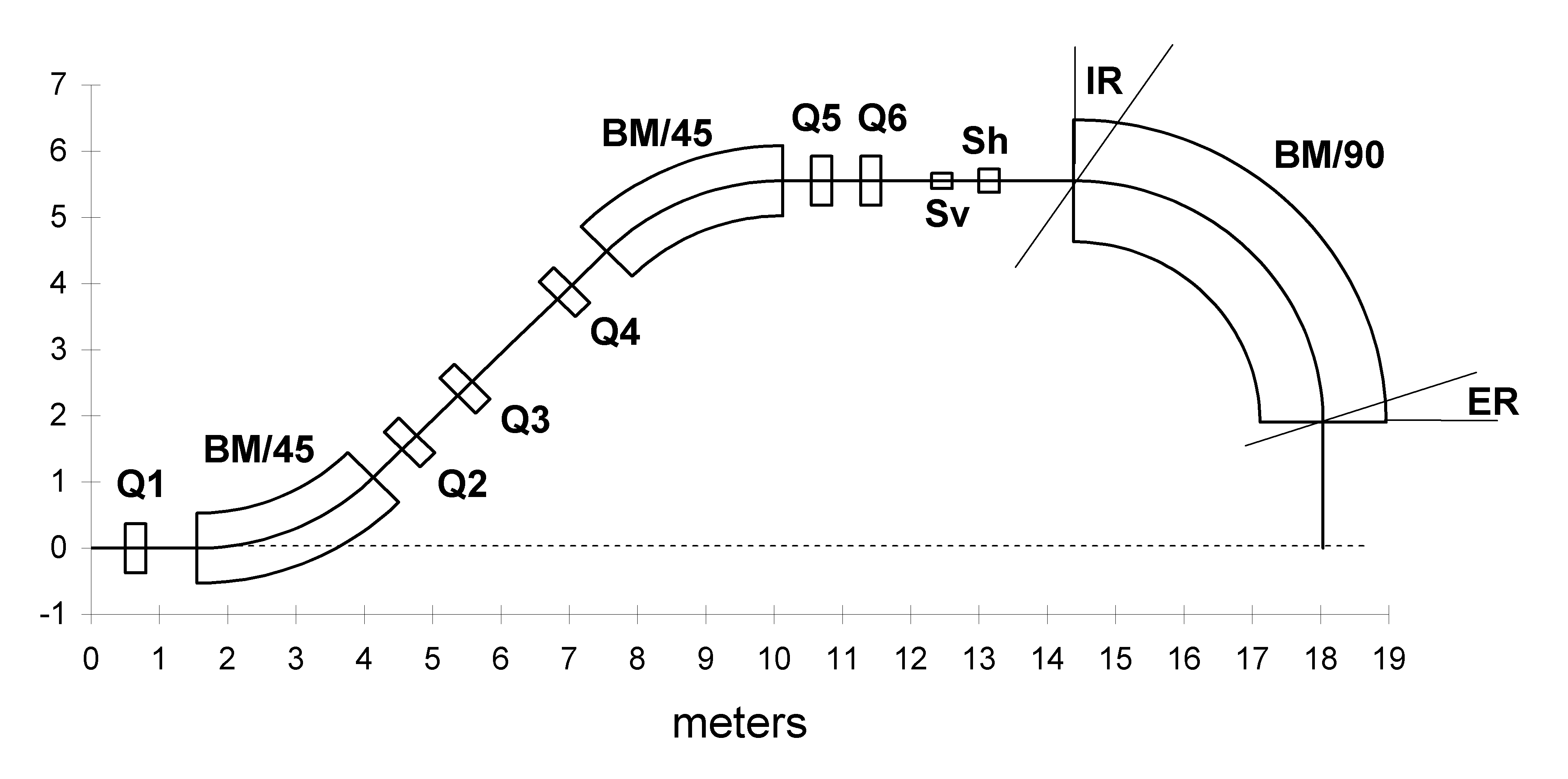}}
\caption{Layout of the ``Pavlovic" gantry with a minimised number of quadrupoles (Q1 to Q6) and three dipoles (BM45 and BM90). Sh and Sv are the horizontal and vertical (respectively) scanning magnets. (courtesy Prof. Marius Pavlovic/STU)\label{pavlovic}}
\end{figure}

A variety of optical schemes exist for downstream scanning, but for upstream scanning the ``Pavlovic" design is the foremost \cite{pavlovic05}.
In the Pavlovic three-dipole design the positioning of both scanning magnets upstream of the final $90 \,^{\circ}$ dipole minimises the gantry radius whilst still providing an infinite SAD; the number of quadrupoles needed to match the beam from the coupling point at the end of the beam transport to the isocentre has been minimised to be as few as six, as shown in Fig.~\ref{pavlovic}.
As well as providing parallel scanning the gantry optics design also manage the dependence of input beam size with gantry angle, which is a more significant issue for the usually non-symmetric beams from synchrotrons, but is solvable \cite{Badano:385378}.

%\caption{Layout of the ``Pavlovic" gantry with a minimised number of quadrupoles (Q1–Q6) and three dipoles (2 \star BM/45 and BM/90). Sh and Sv are the horizontal and vertical (respectively) scanning magnets. (courtesy )\label{pavlovic}}

Besides the minimisation of gantry length and radius, another method of reducing the footprint of a treatment room is to utilise rotations of less than $360 \,^{\circ}$; using $180 \,^{\circ}$ of rotation from vertically-downward to vertically-upward (on one side of the patient), nearly half the building footprint may be saved.
It is then of course necessary to rotate the patient table for some fields; typical isocentric gantry and table rotation speeds are similar at $\approx$1 revolution per minute, so ``$180 \,^{\circ}$'' gantries may not significantly add to treatment time.
However an added problem is introduced, which is to maintain accurate patient position registration after the table rotation.
Nevertheless, the use of restricted gantry rotation can significantly increase the patient treatment capacity on an existing clinical site, or on the restricted space available in the urban population centres often attractive for siting hospital facilities.
A final method which has been considered is to use obliquely-exiting beams from the gantry, which allows for a smaller radius but restricts the range of treatment field directions for supine patients \cite{Pavlovic:1999bm}.

As mentioned earlier, it is also possible to directly mount the accelerator source onto the gantry, as may be done with high-field superconducting synchrocyclotrons; several examples are presently under construction and commissioning by Mevion in the USA.
Gantry sizes are similar to those for normal-conducting gantries fed by external sources and are typically proposed as either single-room or scalable clinical solutions, noting that single-room solutions are also offered which do not mount the source on the gantry.
Another scheme is TULIP, a hybrid method in which a small (e.g. 60~MeV) cyclotron injects protons into a combined linac and gantry.
Again, this is proposed as a single-room solution, but has not yet been implemented although the accelerating structures have been developed as part of the related cyclinac approach described earlier \cite{Degiovanni:2013vi}.

\section{Future Developments}

\subsection{Particle Sources}

The conventional accelerator source technology utilised in current proton and carbon therapy is now rather mature; a good overview of those technology trends has been performed by Amaldi et al. \cite{amaldi10}. Here, we discuss three particular technology developments with recent significant results.

\subsubsection{Dielectric Wall Accelerators}
Dielectric wall accelerators (DWA) for proton therapy are a development from technology originally developed for high-intensity Blumlein-type linear induction accelerators to conduct flash radiography.
Recent work uses improvements in solid state switching technology to achieve direct, cavity-less acceleration with potential gradients as high as 100 MV/m \cite{caporaso08, caporaso11, chen11}.
If achieved in a complete accelerator, such a gradient would allow a complete acceleration system to 250~MeV to be only several metres in length, potentially allowing it to be mounted on a gantry.
Sample structures have been demonstrated and a commercial company (CPAC) is offering a solution, with a prototype system proposed to be ready for clinical testing around 2015.
%It is generally seen as a very promising technology.

\subsubsection{Fixed-Field Alternating Gradient Accelerators (FFAGs)}
As mentioned earlier, one of the disadvantages of the classical cyclotron is the loss of synchronism with a fixed-frequency accelerating cavity as the accelerating protons become more relativistic; the synchrocyclotron alleviates this problem by varying the RF frequency during bunch acceleration, with the concomitant penalty of reduced intensity since only one bunch may be present in the synchrocyclotron per RF frequency sweep.
However, the relativistic limit has limited the practically-achievable proton energies in either method to $\approx$250~MeV.

FFAGs are an adaptation of the cyclotron with significant differences that enable higher energies to be achieved \cite{Trbojevic:2009dc,VerduAndres:2011va}.
In both the cyclotron and the FFAG the magnet field is fixed in time but may vary both azimuthally and radially; the difference is that in the FFAG there is an alternating gradient in successive dipoles such that the circulating particles see strong rather than weak focusing.
The FFAG is thus akin to a strong focusing synchrotron (albeit without varying magnetic fields) and without separated-function magnets (i.e. separate dipoles and quadrupoles) in its layout.

The so-called scaling FFAG is similar to the cyclotron in having a betatron tune that is approximately constant during acceleration, achieved by varying the magnetic field nonlinearly with radius.
The term ``scaling" refers to the fact that the orbit shape scales with energy to maintain a constant betatron tune, so that the bunches may be kept away from damaging resonances (similar to the procedure adopted in most particle accelerator designs).
However, like the cyclotron, the orbit radius varies greatly with energy and requires large-aperture magnets to accommodate a significant energy range, say from the 10s of MeV at injection to the extraction energy for treatment.
The advent of fast swept-frequency accelerating cavities around a decade ago enabled the first demonstration of proton FFAGs that could accelerate particles in less than 1~ms; energies up to 150~MeV have now been demonstrated in Japan as part of the development towards future high-power proton accelerators \cite{Uesugi:2008wz}, and applied at lower energy to the efficient generation of neutrons for boron-neutron capture therapy \cite{Mori:2006ef}.

The non-scaling FFAG uses strong non-linear magnets to allow acceleration over a large energy range whilst retaining small-aperture magnets, and has recently been demonstrated experimentally for electrons \cite{machida12}.
Although the magnets may be made smaller, the lack of orbit scaling with energy inherently gives rise to the crossing of resonances during acceleration; it has been demonstrated that this may be successfully carried out if magnet tolerances are sufficiently well controlled, but a demonstration of the non-scaling principle with protons has yet to be carried out.

Two detailed design studies have been carried out to examine the use of FFAGs for medical therapy, although others have also been performed \cite{Misu:2004dr,Keil:2007ep}.
The first - RACCAM - was a design study for a multi-room treatment centre based on a 180~MeV normal-conducting scaling FFAG \cite{Antoine2009293}.
Energy variation in 250 keV steps (one step per second) is planned at five possible extraction points, each potentially equipped with a patient gantry.
However, the scanning speed is envisaged to be significantly better than that obtainable with present-day synchrotrons.
The second design study - PAMELA \cite{peach10} - proposed a two-ring non-scaling FFAG delivering both protons up to 250~MeV and carbon ions up to 400~MeV/u; a novel superconducting magnet triplet design was developed during the course of the study to allow nonlinear correction of the magnetic lattice \cite{Sheehy:2010kn,Witte:2012gk}.
The proposed advantage of PAMELA is the possibility of having variable energy pulses delivered at rates as high as 1~kHz; the difficulty resides in the need for rather complex superconducting magnets and a very fast pulsed extraction system to produce the rapid energy variation.
Whilst there is potential commercial interest in FFAGs, there are at present no planned clinical centres that utilise them.

\subsubsection{Laser Proton Acceleration}

Protons and other ions may be accelerated using laser pulses in several ways; recent reviews have been given by Daido et al. \cite{Daido:2012fl}, Norreys~\cite{Norreys:2011kg} and Macchi et al. \cite{Macchi:2013kc}.
One recent advance has been the demonstration of so-called target normal sheath acceleration (TNSA), whereby protons are accelerated from the rear of a thin target illuminated by a strong laser pulse due to the electron pressure within the target \cite{Snavely:2000ew}.
Acceleration of protons and ions has been demonstrated up to 10s of MeV with reasonable beam quality and energy spread \cite{Daido:2012fl,Hegelich:2006bt,Metzkes:2011jg}; however, whilst scaling to clinically-relevant energies up to 250~MeV has been modelled there has not yet been demonstration to those energies.
A number of research groups are engaged in achieving this goal, and are considering how to deliver the pulse repetition rates required for clinical application \cite{cowan10} and to utilise the resulting protons \cite{Richter:2011dp}.
One avenue is the use of radiation-pressure (``light sail") acceleration \cite{Robinson:2008fj}, in which the incident photons themselves impart momentum to the accelerated ions.
%to improve on the pulse repetition rate and energy spread such that the proton bunches could be used for therapeutic purposes.
Whilst the use of laser technology is very promising, it is likely that it will be a few years before clinical experiments are carried out and there are no commercial companies with plans to offer laser-accelerated protons for clinical use.
One possible advantage of laser-based acceleration, like other compact proton sources, is the ability to mount the entire particle source onto a delivery gantry.
Efficient collection and focusing of these laser-derived particles requires a compact delivery system, based on either conventional magnets (either quadrupolar or solenoidal) \cite{Ma:el,Hofmann:2013gm} or on Gabor lenses \cite{Pozimski:2005eh,Pozimski:2013vo}.

\subsection{Gantries}

Similar to accelerator sources, delivery gantries have seen a great deal of development; in contrast though, gantry design has evolved towards one of two canonical design approaches, either an isocentric gantry delivering externally-generated protons, or a gantry-mounted source delivering protons direct to the patient. Here, we discuss two approaches for making isocentric gantries more compact.

\subsubsection{Superconducting Magnets}

All three of the driving forces in gantry development, namely:
\begin{itemize}
  \item Reducing their size and cost;
  \item Increasing the beam energy for the same size;
  \item Utilising ions heavier than protons;
\end{itemize}
require increased magnetic field strength (both for bending and focusing) integrated over the beam path.
Of course, the gantry size is a significant cost driver of the whole treatment facility; a gantry treatment room is comparable in size or larger than most accelerator sources.
Also, gantries for carbon ion therapy require larger magnets and in order to have a wider adoption of carbon-ion therapy it will be necessary to reduce the size and mass of the magnets.
One method to achieve this is to adopt superconducting technology.

The advantages of superconducting magnets are that they can support a much larger magnetic field than a normal-conducting magnet and because there are almost no resistive losses in the magnet circuit they cost less to power.
Widely used in particle accelerators used for research for many years, they consist of either iron-cored or core-less magnets incorporating superconducting coils which are typically cooled to less than 4 K using liquid helium; the use of high-temperature superconductors in this application has not yet been widely demonstrated.

\begin{figure}[htb]
\centerline{\includegraphics[width=0.8\textwidth]{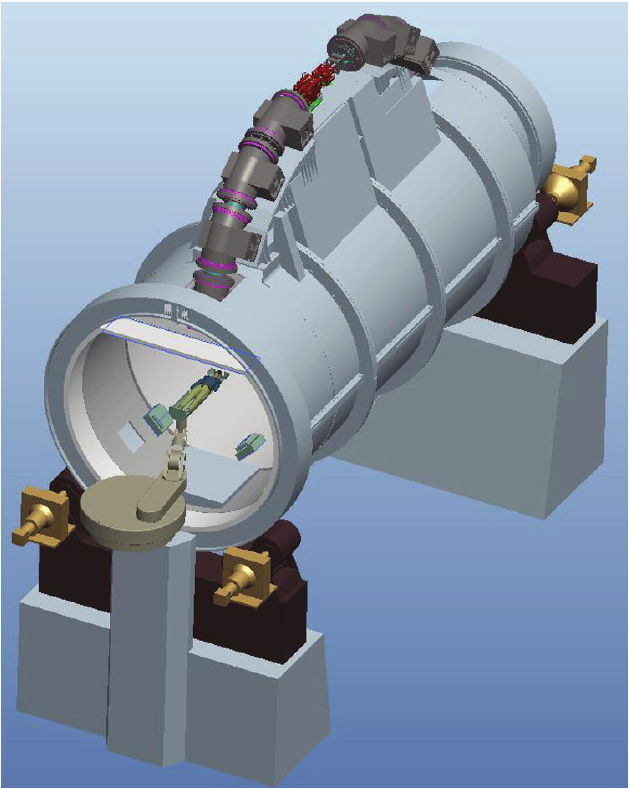}}
\caption{Three-dimensional image of the NIRS superconducting rotating gantry for heavy-ion therapy. (courtesy Dr. Yoshiyuki Iwata/NIRS) \label{NIRS_schematic}}
\end{figure}

\begin{figure}[htb]
\centerline{\includegraphics[width=1.0\textwidth]{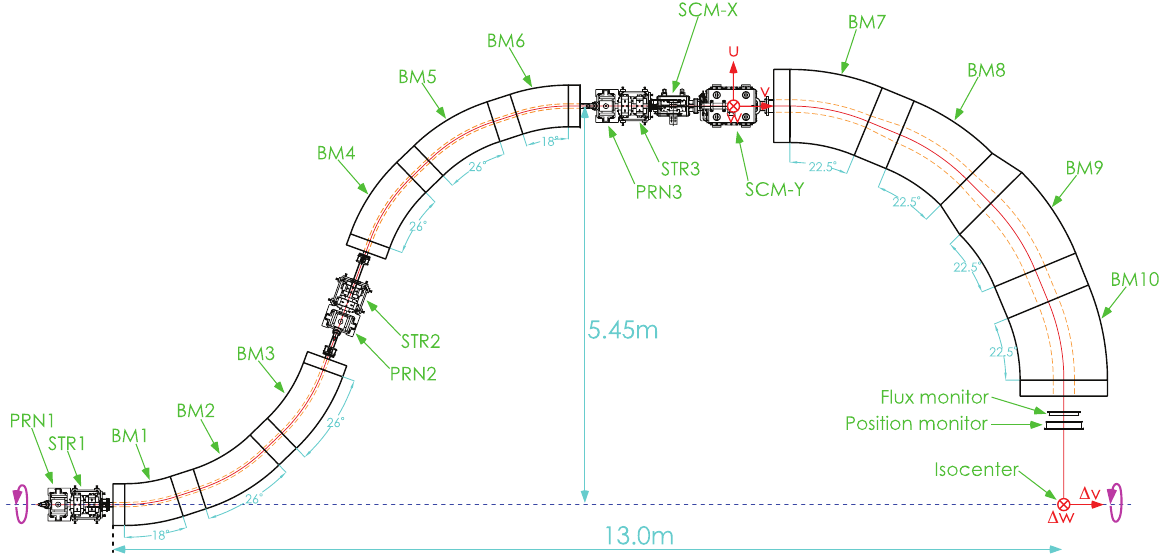}}
\caption{Layout of the NIRS superconducting rotating gantry. The gantry consists of ten superconducting magnets (BM1-10), a pair of scanning magnets (SCM-X and SCM-Y) and three pairs of beam profile-monitor and steering magnets (STR1-3 and PRN1-3). (courtesy Dr. Yoshiyuki Iwata/NIRS) \label{NIRS_layout}}
\end{figure}

Several groups - some with commercial involvement - have proposed core-less curved superconducting dipoles which produce fields up to 3.3~T or more \cite{robin11}.
Core-less superconducting dipole magnets have been prototyped to some extent, but suitable magnetic field distribution and quality have not been demonstrated experimentally for all the magnet types required in a practical gantry.
Superconducting gantry designs are underway at ETOILE \cite{bajard08} and NIRS \cite{iwata12}; for example, the 3.3~T field proposed for ETOILE significantly reduces the dipole bend radius for 425~MeV/u carbon ions to around 2~m, resulting in gantry dimensions (13.5~m long by 4~m radius) and mass (210~tonnes) which are comparable to normal-conducting gantries for protons. An example of a superconducting gantry design for NIRS is shown in Fig.~\ref{NIRS_schematic} and Fig.~\ref{NIRS_layout}.
The NIRS design envisages their synchrotron producing carbon ions between 430~MeV/u and 56~MeV/u in 200 steps, corresponding to a 1~mm to 2~mm range in water \cite{iwata10}.
Ridge filters are then used to produce a mini-SOBP of between 1 to 3~mm.
Similarly, the superconducting magnets will change their field to match the beam energy within 200 ms.
The results of tests of the prototype magnets and cryostats are promising \cite{Iwata2013}.

Although there are savings in the electrical power circuits that supply superconducting magnets coils when compared to normal conducting coils, superconducting magnets are more problematic to use than normal-conducting ones as they must be cooled to a few kelvin.
There is also the higher capital cost associated with the magnets themselves and for their cooling system.
The use of conventional liquid helium-filled cryostats is probably not possible as liquid movement during rotation would lead to potential quenching, and so the approaches described above typically plan to use cryocoolers.
These are cryogen-free heat pumps \cite{Radebaugh:2009bo} and so allow rotation of a magnet on a gantry, but they typically have a more limited capacity than cryogenic-liquid cooling systems.
The NIRS design envisages a pre-cooling procedure using liquid nitrogen to reduce the temperature to $\approx$70~K, prior to the use of cryocoolers to further cool the magnets to 4~K.
This will reduce the total time to get from room temperature to 4~K from about a month to six or seven days.
A further complication is that superconducting magnets are prone to quenching when new (and until a period of training has been undertaken) and also when required to change field rapidly.
Tests at NIRS have shown that quenches can be recovered from immediately; although realistically it may take up to two hours.
It is not yet clear how quenching would be managed if it happened part-way through a treatment fraction.
That being said, significant progress has been made in Japan toward the realisation of a superconducting gantry and it is likely that one will be operational soon.

%This is likely to be unacceptable for clinical operation.
%Although it is possible to change the magnet field at a restricted rate (a about 1 second to accommodate a 1 to 2$\,\%$ momentum change), superconducting gantries are not ideally suited to treating successive layers of the treatment volume with a progressive changes in beam energy.ADD INFO from Iwata

\subsubsection{FFAG Gantries}

\begin{figure}[htb]
\centerline{\includegraphics[width=1.0\textwidth]{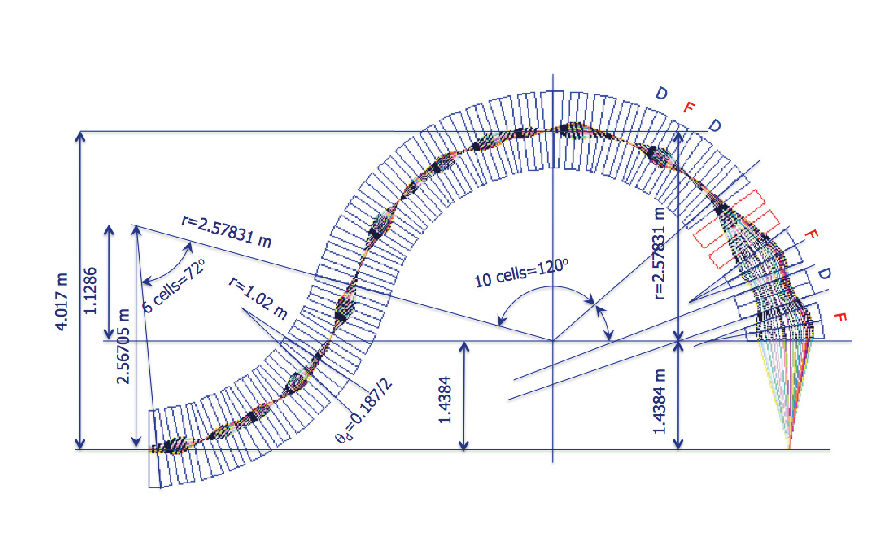}}
\caption{Schematic layout of proposed superconducting FFAG isocentric gantry for carbon therapy \cite{trbojevic11}, utilising triplet magnets in which the defocusing dipole has a maximum field of 5.58~T and a gradient of of -93~T/m. This design allows a carbon gantry in principle to be as small as existing proton gantries . (courtesy Dr. Dejan Trbojevic/Brookhaven National Laboratory) \label{ffaggantry}}
\end{figure}

An alternative approach to conventional beam optics using superconducting magnets is the use of fixed-field, alternating-gradient (FFAG) optics which in principle give very large energy acceptance with a small magnet aperture.
FFAGs gantries make use of combined-function dipole magnets with large focusing gradients that alternate in sign along the beam path.
The large gradients, and the use of non-scaling beam optics design, minimise the dipole-generated beam dispersion and thereby restrict the aperture required for a given energy range; energy scanning may then be carried out without varying the magnetic field of the gantry magnets.
Thus some designs propose the use of permanent magnets to reduce the weight.
FFAG gantries may also utilise superconducting or conventional normal-conducting magnet technology, and a number of optical and technological solutions have been proposed \cite{Trbojevic:2007jk,Keil:2007ep,trbojevic11}.
It is proposed in particular that FFAG gantries for carbon ion transport may be significantly smaller and lighter than other designs.
An example layout of a superconducting FFAG gantry is shown in Fig.~\ref{ffaggantry}.
In treatment situations where the energy range that is required is larger than is available from the FFAG gantry optics, it is proposed to use a small number of magnetic field settings to cover the depth range.
Although a number of innovations have been proposed, including the provision of parallel scanning mentioned previously, an FFAG gantry has not yet been demonstrated.

%\subsection{Technology Trends in Treatment Planning and Verification}

\section{Summary}

%\subsection{Particle Sources}

Following a long period of experimentation before becoming fully accepted, proton and other light ion beam therapies are now well-established clinical techniques.
Whilst IMRT may give excellent dose conformation to the tumour, protons provide the possibility of significantly reducing the dose to surrounding tissues and organs at risk.
For the treatment of particular types of cancer this advantage outweighs the additional technology costs, particularly in paediatric treatments where the induction of secondary tumours or other side-effects can have particularly deleterious impacts on the patient in later life.

In recent years - and particularly in the last decade - proton treatment technology has advanced greatly, and the advent of scanned beams and IMPT have significantly improved the capability of proton treatment.
Mature technology solutions are now available in the marketplace, but there remain opportunities for improvement, either to further improve clinical capability or to reduce cost; both of these aims are important to realise the full potential of proton and ion therapy.
There is already a well-established pattern in this field, in which accelerator technology research from state-funded research (e.g. at national laboratories) has been successfully transferred and developed to clinical practice by commercial organisations.
Ion beam therapy has benefitted from the forward-looking nature of many companies in the field and from the investment they have been brought to create new products.

A number of novel accelerator technologies are being developed that aspire to provide greater capability or reduced cost, but will have yet to show their advantages in comparison to existing offerings.
At the accelerator source end, we note the steadily increasing use of superconducting technology for proton acceleration, which has aimed to reduce cost and size, enable mobile sources, or enable higher energies to be achieved.
We also note dielectric wall accelerators and non-scaling FFAGs as being disruptive technologies: they promise significantly improved capabilities (DWAs offering much reduced size and ns-FFAGs promising more rapidly variable energy), but neither have yet been demonstrated in a clinical setting.

Carbon-ion therapy has become the de-facto standard for (non-proton) ion therapy and makes inherently greater demands on the technology compared to protons.
As a result, it has pushed the development of higher-gradient acceleration and higher magnetic fields for beam delivery.
Several groups have been involved in developing superconducting magnets suitable for gantries and we believe this is a key technology for the near future.
The strong focusing and compact magnet arrangements available in FFAG gantries could also offer potential benefits for scanning and to obtain smaller size, but given the more complex beam optics it will be necessary to construct a prototype.

It is not only carbon ions that may benefit from the use of higher fields or other methods to achieve smaller gantry sizes than the present successful but conservative designs.
In proton therapy, there is a desire to reduce the gantry size to enable a larger number of treatment rooms on a given site or to enable higher energies to be transported, e.g. to enable the use of proton computed tomography. Given the typically urban location of treatment centres this desire will persist.
We note here the trend towards opting for reduced gantry angle range ($\approx180 \,^{\circ}$ instead of $\approx360 \,^{\circ}$) for which there is a tradeoff between capital cost and throughput.
Throughput remains a key indirect cost driver and the possible reliability and throughput advantages of gantry-mounted sources has only begun to be addressed by the initial example developed by Mevion.

In conclusion, we believe that despite the great advances in ion therapy technology in the last few years, there remain significant opportunities.
In the near-term we believe superconducting technology will play an increasingly important role, particularly in beam gantries.
Further on, new compact accelerator sources - possibly including laser-based acceleration - offer the ability to provide gantry-mounted solutions if sufficient gradient can be achieved; these could potentially offer throughput or cost advantages, but have to compete against the established mature solutions based on conventional accelerator technology.

\section*{Acknowledgments}

The authors would like in particular to thank Dr. Yoshiyuki Iwata and Prof. Marius Pavlovic for useful discussions and information.

This work was supported in part by the United Kingdom Science and Technology Facilities Council.

%This section should come before the References. Dedications and funding information may also be included here.
%\subsection{Treatment Planning}

\bibliographystyle{unsrt}

\end{document}